\documentclass[letterpaper, 10pt, conference]{ieeeconf}
\IEEEoverridecommandlockouts   
\usepackage{cite}
\usepackage{amsmath,amssymb,amsfonts}
\usepackage{graphicx}
\usepackage{textcomp}
\usepackage{xcolor}
\usepackage{hyperref}
\usepackage{makecell}
\usepackage{booktabs}
\usepackage{adjustbox}
\usepackage{placeins}
\usepackage{multirow}
\usepackage{tabularx} 

\title{Frugal Collective Perception: Context-Aware Adaptive Reporting for Safety-Critical C-ITS}

\author{Romain Tessier\,$^{1,2}$, Bruno Monsuez$^{1}$, Adriana Tapus$^{1}$, Oyunchimeg Shagdar$^{2}$
\thanks{$^{1}$\textit{U2IS, ENSTA, Institut Polytechnique de Paris}, 828 Bd. des Maréchaux, 91120 Palaiseau, France. \texttt{\{romain.tessier, bruno.monsuez, adriana.tapus\}@ensta.fr}}%
\thanks{$^{2}$\textit{Ampere Software Technology}, 1 Av. du Golf, 78280 Guyancourt, France. \texttt{oyunchimeg.shagdar@ampere.cars}}%
}

\begin{document}

\maketitle

\begin{abstract} 
Ensuring safety and scalability in Collective Perception Service (CPS) remains a key challenge for Cooperative Intelligent Transport Systems (C-ITS). Conventional CPS enhances perception by broadcasting Collective Perception Messages (CPMs). However, its reliance on transmitting a potentially large volume of context-irrelevant information at high frequency leads to network congestion, processing delays, and poor scalability. We propose a Context-Aware Adaptive Filter that dynamically adjusts CPM content and transmission frequency based on contextual relevance and situational criticality. By prioritizing safety-critical objects and interactions, the proposed approach prevents information overload while preserving timely updates for decision-making. An end-to-end SUMO–Artery simulation evaluates safety, decision-making efficiency, and communication cost under different computational capacity tiers and transmission rates. Results show that the proposed adaptive filtering mechanism achieves safety performance comparable to conventional CPMs transmitted at the maximum allowed frequency (10~Hz), while reducing communication volume by over 93\% and preventing queue saturation. This demonstrates that context-aware adaptivity enables CPS to remain both scalable and safety-compliant across heterogeneous computing platforms.
\end{abstract}


\section{Introduction}
Ensuring safety and efficiency in increasingly complex traffic environments remains one of the major challenges for automated driving systems. While modern vehicles are equipped with advanced sensors such as LiDAR, radar, and cameras, their perception remains fundamentally constrained by occlusions, limited sensing range, and adverse environmental conditions. Cooperative Intelligent Transport Systems (C-ITS) have emerged as a promising solution by enabling vehicles to exchange information and extend their situational awareness beyond the line of sight. Within this framework, the Collective Perception Service (CPS) plays a central role by allowing vehicles to share information about locally detected objects through standardized Collective Perception Messages (CPMs). Such collaboration strengthens the capability of connected vehicles to predict hazards, identify obscured road users, and execute safer decisions in real time.

Despite these advantages, the scalability of CPS is challenged by the exponential growth of shared information as the number of connected participants increases. Each CPM contains descriptions of multiple detected objects, and broadcasting them at high frequency can quickly lead to network congestion, increased latency, and excessive computational overhead at the receiver side~\cite{schiegg_collective_2020}. 
Furthermore, receivers cannot determine the usefulness of shared information before processing it, which leads to wasted resources and limits the scalability of the system~\cite{huang_select2drive_2025}. 
This issue is particularly critical in late-fusion collaborative perception: since the emitter has already processed its local observations, it is in a better position to assess their potential relevance before transmission. Optimizing CPS thus requires moving from indiscriminate broadcasting to emitter-side strategies that regulate what is shared and at which rate~\cite{lyu_survey_2025}.

The key challenge in C-ITS environments is that participants and contexts change rapidly, preventing an emitter from having complete knowledge of a receiver’s state, control objectives, or situational awareness. Therefore, relevance must be inferred from the observable situation and the expected impact of sharing a given object. To address this, we argue for a dual-rate communication mechanism: all detected objects should be transmitted at a low baseline frequency to ensure that peers maintain a minimum level of scene completeness, while objects with higher contextual value, such as those involved in potential conflicts, occlusions, or safety-critical interactions, should be transmitted more frequently. This strategy balances communication efficiency with safety relevance by reducing redundant data while preserving timely updates for critical information.

In this work, we explore the optimization of such an adaptive filtering mechanism for CPMs. Building on recent advances in semantic communication and context-aware collaborative perception, we design a framework in which emitters dynamically adjust both CPM content and frequency based on the contextual value of the detected objects. By aligning message generation with decision-making importance, our approach reduces network overhead, preserves computational feasibility, and ensures reliable delivery of safety-critical information in real-time.


\section{Related Work}
Collective perception has been extensively studied in recent years, with much attention devoted to intermediate fusion, where agents exchange feature representations instead of raw sensor data or final object lists. This design aims to balance perception accuracy with communication efficiency. When2com~\cite{liu_when2com_2020} introduced a grouping strategy to decide \textit{when} and \textit{with whom} to share, while V2X-ViT~\cite{xu_v2x-vit_2022} leverages transformers to integrate heterogeneous features and address asynchrony, pose errors, and variability. Where2comm~\cite{hu_where2comm_2022} proposed spatial confidence maps to transmit only critical regions, reducing bandwidth by orders of magnitude. PragComm~\cite{huang_select2drive_2025} further improves efficiency by adaptively selecting beneficial features, and Direct-CP~\cite{tao_direct-cp_2024} demonstrates that end-to-end pipelines can achieve effective fusion without engineered communication policies. While these works show that adaptive feature-level fusion enhances collaborative perception, exchanging high-dimensional features—whether sparsified or attention-guided—still entails considerable overhead and often requires tightly coupled architectures, limiting interoperability in heterogeneous V2X systems.

Another important direction focuses on task-oriented collective perception, where communication and perception are optimized for downstream objectives rather than generic scene understanding. Fang et al.~\cite{fang_r-acp_2025} introduced R-ACP, a robust framework that combines self-calibration and feature sharing under an information-bottleneck principle, explicitly modeling timeliness with the ``Age of Perceived Targets.'' Similar strategies appear in multi-robot perception, where active sensing transmits only task-relevant information to support navigation or manipulation~\cite{singh_multi-agent_2024}. Other works adopt task-agnostic approaches, such as collaborative scene completion or semantic reconstruction~\cite{li_multi-robot_nodate, gao_stamp_2025}.

Semantic representations have also been explored to guide perception and communication. Metric-semantic maps and scene graphs provide compact, queryable structures that can be exchanged to support decision making~\cite{garg_semantics_2020}. Kimera-Multi~\cite{rosinol_kimera_2021} showed how distributed semantic scene graphs can be built in real-time for bandwidth-efficient multi-robot collaboration. The survey \textit{From SLAM to Situational Awareness}~\cite{bavle_slam_2023} traces the shift from low-level mapping toward semantically enriched situational awareness, underscoring the importance of reasoning at the object level for intelligent transportation systems.

Most recently, Lusvarghi et al.~\cite{lusvarghi_search_2025} proposed a context-aware paradigm for semantic and task-oriented V2X communications, where message curation is guided by the \textit{relevance} of information to the receiver’s driving task. Their analysis shows that relevance-aware filtering can substantially reduce network load while ensuring safety-critical information is delivered. However, this vision remains largely conceptual and does not specify how relevance is calculated. This gap motivates our approach, which makes relevance explicitly computable through semantic maps and reasoning.

In contrast to prior work, our approach explores a frugal late-fusion paradigm grounded in the ETSI CPM standard. By leveraging a semantic map to predict the contextual relevance of detected objects for other connected agents, the CPM emitter can adaptively filter its outputs and prioritizes safety-critical information. This strategy minimizes communication overhead while directly supporting downstream applications.

\section{Problem Statement}

Collective Perception Services (CPS) promise safer driving through the exchange of object-level information. However, their effectiveness is constrained not only by communication limits but also by the computational demands placed on receiving vehicles. The following subsections examine these constraints and introduce a processing model to better capture their impact.

\subsection{Challenges in CPM Transmission}

Each Collective Perception Message (CPM) contains a header and a set of \textit{Perceived Object} Containers that describe detected objects. These messages are broadcast periodically, typically between 1~Hz and 10~Hz depending on traffic dynamics and channel conditions, and convey processed object-level information rather than raw sensor data. Transmitting vehicles aggregate their local detections into object hypotheses, while receivers fuse incoming CPMs with their own perception using spatial, temporal, and confidence-based matching~\cite{lyu_survey_2025}.

ETSI standards provide reporting rules to reduce redundancy, including the prioritization of vulnerable road users (VRUs), updates on significant state changes, and suppression of stable objects under congestion~\cite{lyu_survey_2025, noauthor_ts_nodate}. While effective in limiting channel load, these rules overlook the computational burden imposed on receiving vehicles. In practice, receivers must decode and fuse potentially large numbers of objects, which may exceed their real-time processing capacity. Thus, the challenge in Collective Perception Services (CPS) is not only to limit communication overhead but also to ensure that shared information remains usable within processing deadlines.

\subsection{Proposed Perception Processing Model}

To analyze the impact of collective perception on real-time decision-making, we introduce a simplified perception processing model. The objective of this model is not to provide hardware-accurate timing predictions, but to capture the relative scaling behavior of decoding and fusion operations as the number of perceived objects increases. By explicitly modeling how computational load grows with scene complexity and incoming CPM volume, the model allows us to study when and why conventional CPS strategies become impractical under limited processing budgets. 
Incoming CPMs from surrounding agents are buffered before being processed, and the perception cycle is bounded by a maximum period ($T_{total}$). Local perception is assumed to rely on an image analysis pipeline, which introduces a fixed delay   ($\alpha$) independent of the number of detected objects. By contrast, decoding CPMs requires parsing each reported object, such that decoding \(M\) \textit{Perceived Objects} takes  
\begin{equation}
T_{decode}(M) = \gamma + \delta M,
\end{equation}
where \(\gamma\) is the fixed decoding overhead and \(\delta\) the per-object decoding delay. 
Fusing \(N\) local detections with \(M\) decoded \textit{Perceived Objects} introduces  
\begin{equation}
T_{fusion}(N,M) = \epsilon + \zeta N M,
\end{equation}
where \(\zeta\) represents the per-pair fusion delay. The overall cycle time is as follows:  
\begin{equation}
T_{\text{cycle}} = \max \big( \alpha, T_{\text{decode}}(M) \big) + T_{\text{fusion}}(N,M),
\end{equation}
subject to the real-time constraint \(T_{\text{cycle}} \leq T_{\text{total}}\).  

This model highlights that the number of \textit{Perceived Object} that can be processed per cycle depends on the local scene complexity. When \(N\) is small, the system is decode-limited and can accommodate more CP data. As \(N\) increases, the system becomes fusion-limited, and incoming objects must be selectively restricted.

\section{Context-Aware Adaptive Filtering}
Collective perception requires mechanisms that regulate which objects are shared, when they are transmitted, and at what frequency. Static reporting rules, as defined in current standards, are insufficient to adapt to varying traffic densities, communication load, and safety-critical contexts. To address this limitation, we propose a Context-Aware Adaptive Filtering strategy, in which CPM generation is driven by the contextual relevance of perceived objects rather than by fixed-rate broadcasting rules. By aligning reporting policies with situational criticality and expected decision impact, the proposed approach reduces redundant communication while ensuring that safety-critical information is delivered in a timely and reliable manner.

\subsection{Traffic Scene Representation}

\begin{figure} [h!]
    \centering
    \includegraphics[width=0.95\linewidth]{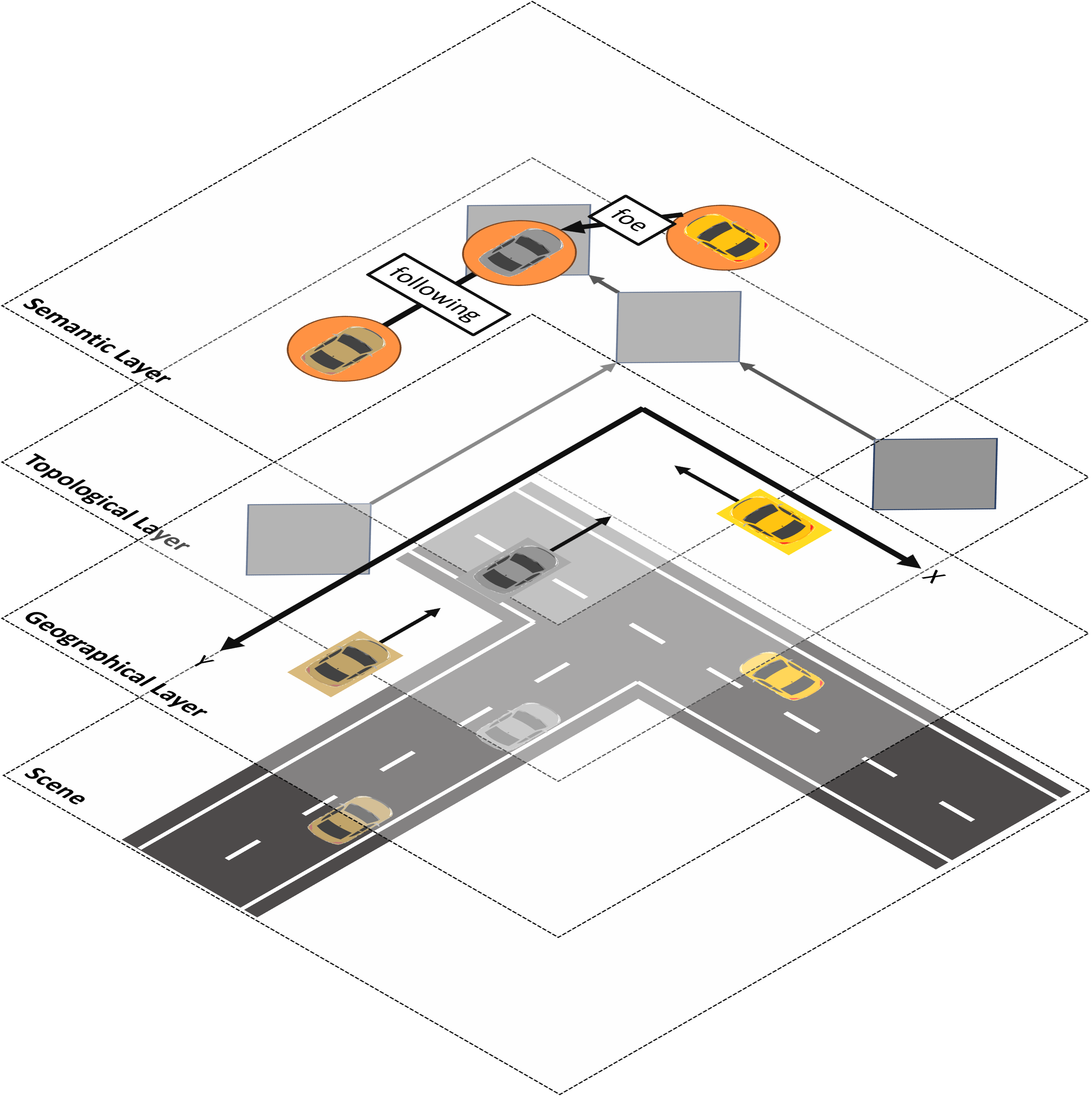}
    \caption{Three-layer traffic scene representation.}
    \label{fig:traffic}
\end{figure}

To support context-aware filtering and situational reasoning, we introduce a three-layer traffic scene representation, illustrated in \autoref{fig:traffic}. This hierarchical structure organizes environmental knowledge from raw kinematics to high-level interactions, enabling the estimation of contextual relevance and safety-critical relationships among road users.

The \textit{geographical layer} encodes the kinematic state of each detected object in a global $(X, Y)$ coordinate system, including position, velocity, acceleration, and short-term motion predictions.  
The \textit{topological layer} represents the road network as a graph, where lanes are modeled as nodes and edges describe their connectivity (e.g., merging, diverging, or crossing). Detected vehicles are mapped onto their respective lanes, linking their motion to infrastructure constraints.  
Finally, the \textit{semantic layer} captures interactions among road users through a relation graph. Two relations are explicitly modeled, as they provide relevant information for inferring conflicts between road users: (i) \textit{follower–leader}, and (ii) \textit{foe}, the latter representing right-of-way conflicts at unsignalized intersections.
\subsection{Adaptive Collective Perception Service}

The proposed context-aware filtering mechanism regulates CPM content based on the contextual relevance of perceived objects. Object reporting is driven by situational criticality, such as the likelihood of conflict at intersections or right-of-way violations, rather than by raw detection frequency. Highly critical objects are always reported, medium-criticality objects are transmitted upon significant state changes, and low-criticality objects are included only when major updates occur. The definitions of what constitutes a ``major'' or ``significant'' change are configurable, allowing the system to adapt to traffic conditions and application requirements.

Reporting is further conditioned on the presence of relevant cooperative agents, ensuring that transmissions remain useful for downstream perception and decision-making. Each object is tracked using a global identifier and assigned a dynamic reporting frequency that depends on both its contextual importance and the time elapsed since its last transmission. This adaptive reporting strategy maintains timely situational awareness while limiting communication load and respecting the processing constraints of receiving vehicles. Redundancy is further reduced by suppressing transmissions when the same object has already been recently reported by another agent and remains sufficiently fresh for safe decision-making.

\subsection{End-to-End Implementation}

To evaluate the proposed approach, we developed an end-to-end simulation framework that tightly couples traffic dynamics with vehicular communication. SUMO serves as the microscopic traffic simulator, providing realistic vehicle trajectories, road topology, and intersection behavior. On top of this traffic simulation, Artery models the communication stack and sensor abstraction. It supports ETSI ITS-G5 message generation and exchange, enabling vehicles to broadcast and receive CPMs. Artery also incorporates a configurable sensor model that emulates onboard perception (e.g., detection range and field of view), which serves as input for collective perception. This integrated setup allows vehicles not only to sense their environment, but also to share object-level information, reproducing realistic cooperative perception scenarios.

Building on this perceptual and communication foundation, the Control module governs each vehicle's decision-making and motion planning. Implemented as a hierarchical finite state machine (FSM), it manages longitudinal behavior according to context-dependent policies within a continuous perception–decision–action loop. The FSM includes five states: Free Drive, Following, Intersection Approach, Wait at Intersection, and Emergency Braking. State transitions and acceleration commands are triggered by fused perceptual inputs that combine geometric, topological, and semantic information, ensuring that vehicle behavior reflects both local observations and cooperative information from the network.

\section{Experiments}
\subsection{Scenario}
The simulation scenario is based on a six-lane two-way arterial road segment with a central unsignalized intersection. The network consists of two main corridors running north–south and east–west, each modeled with three lanes per direction, allowing for a variety of maneuvers including straight-through, left-turn, and right-turn movements. At the center of the scenario lies junction, defined as a priority-controlled intersection, where vehicles approaching from the four directions must negotiate right-of-way rules. This configuration introduces natural conflict points between traffic streams, particularly at left-turns and crossing maneuvers, making it well suited for evaluating cooperative perception in complex interactions.


To populate the network with realistic traffic, we defined multiple vehicle flows covering all major movements through the intersection. Variability in departure speeds was introduced to generate heterogeneous driving behaviors. Vehicles were inserted along 12 different routes, with flows scheduled to begin at staggered times and varying intensities, ensuring a continuous stream of interactions at the central intersection. A total of 66 vehicles were inserted over 200 seconds, corresponding to a traffic intensity of approximately 1,200 vehicles per hour. 

To capture safety-critical interactions, each vehicle was equipped with a Surrogate Safety Measures (SSM) device. SSMs compute potential conflicts by analyzing vehicle trajectories. In our simulations, they recorded metrics such as Time-To-Collision (TTC), which measures the time remaining before a collision would occur if vehicles maintain their current trajectories, and Post-Encroachment Time (PET), which measures the temporal separation between vehicles passing through the same point in space.

To emulate heterogeneous processing capabilities (\autoref{tab:perception_timing}), we consider three representative system tiers inspired by commonly used embedded platforms: Jetson TX2 (low-end), Xavier NX (mid-end), and AGX Xavier (high-end). These tiers capture the diversity of computational resources typically encountered in automotive and mobile robotics systems. The parameter $\alpha$, which represents local perception cost, is derived from the reported inference rates of Complex-YOLOv4 (tiny) on the KITTI dataset~\cite{choe_run_nodate}. The non-local processing parameters ($\gamma$, $\delta$, $\epsilon$, $\zeta$) are chosen to reflect the dominant scaling behavior of CPM decoding and fusion operations: fixed costs for protocol parsing and fusion setup, linear growth with the number of received objects, and bilinear growth during data association. Their values are intentionally set so that fusion becomes the dominant bottleneck as scene and CPM complexity increase, thereby exposing the processing saturation regime targeted by the proposed adaptive filtering strategy. 

\begin{table}[h!]
\centering
\scriptsize
\renewcommand{\arraystretch}{1.4} 
\caption{Perception process parameters for different system tiers.}
\label{tab:perception_timing}
\begin{adjustbox}{width=1\linewidth}
\begin{tabular}{@{}l m{2cm} m{1.5cm} m{1.5cm} m{1.5cm}@{}}
\toprule
\textbf{Symbol} & \textbf{Description} & \textbf{Low-end} & \textbf{Mid-end} & \textbf{High-end} \\
\midrule
$T_{\text{total}}$ & Total perception cycle time   & 150 ms & 100 ms & 100 ms \\
$\alpha$           & Local processing time         & 82 ms  & 60 ms  & 37.5 ms \\
$\gamma$           & CPM decoding time             & 10 ms  & 4 ms   & 3 ms \\
$\delta$           & Per-object decoding cost      & 0.8 ms/object & 0.1 ms/object & 0.08 ms/object \\
$\epsilon$         & Fusion setup time             & 6 ms   & 4 ms   & 1.5 ms \\
$\zeta$            & Per-pair fusion cost          & 0.15 ms/pair & 0.06 ms/pair & 0.015 ms/pair \\
\bottomrule
\end{tabular}
\end{adjustbox}
\end{table}

\subsection{Results}

\begin{figure}[h!]
\centering
\includegraphics[width=0.9\linewidth]{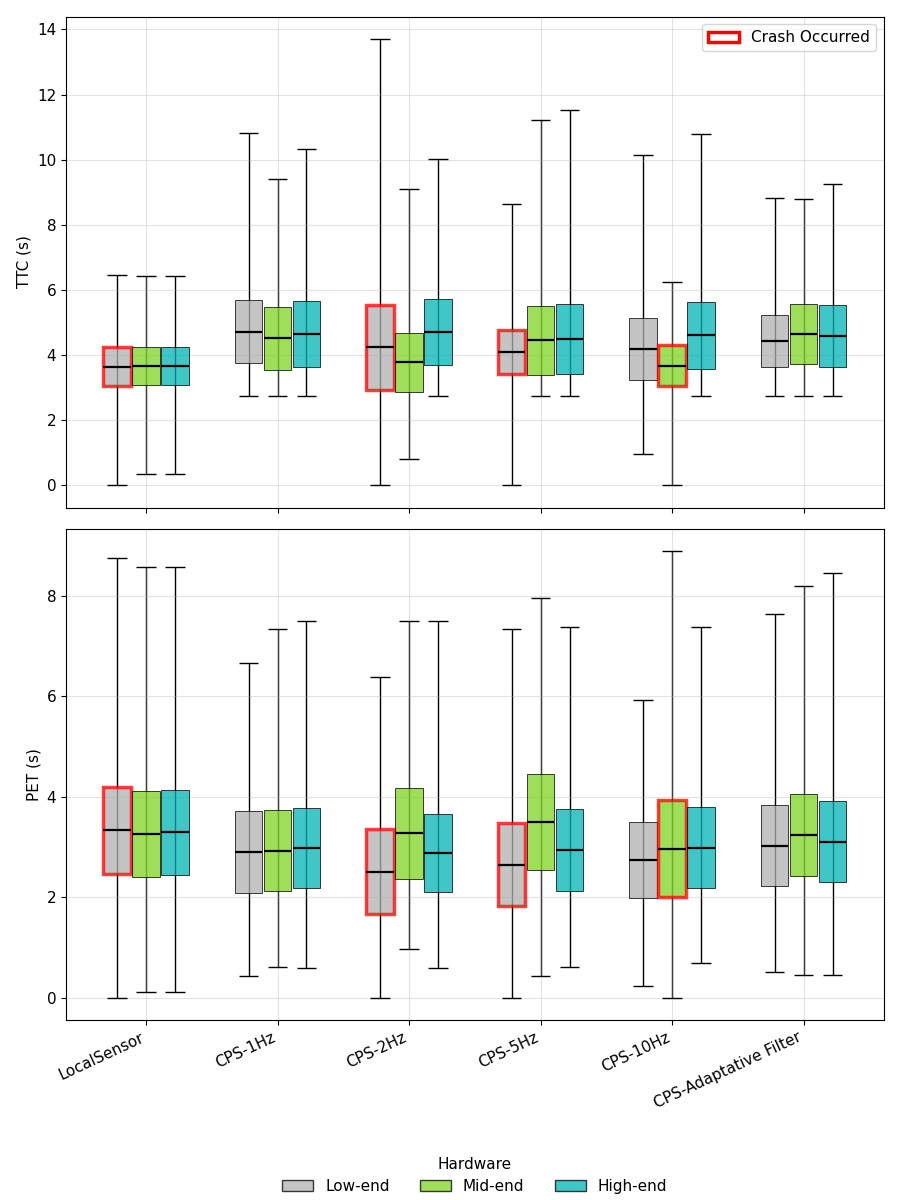}
\caption{Minimum TTC and PET distributions in conflicts across controllers and hardware tiers (Crashes in Red)}
\label{fig:safety}
\end{figure}

\paragraph{Safety analysis}
\autoref{fig:safety} compares the safety indicators \textit{Time-to-Collision (TTC)} and \textit{Post-Encroachment Time (PET)} across different driving controllers and hardware configurations. The \textit{LocalSensor} configuration, which does not support the ability to receive and process CPM, produces much wider distributions, including crashes. This behavior can be attributed to the restricted sensing horizon: without collaborative information, vehicles cannot anticipate hazards hidden by occlusions, leading to abrupt maneuvers and unsafe PET values that sometimes collapse toward zero. Hence, relying solely on local sensing reduces robustness and increases crash likelihood.

The introduction of \textit{CPS} significantly improves both TTC and PET by extending the perception horizon. Across frequencies from $1$Hz to $10$Hz, CPS-equipped vehicles maintain higher TTC values and more stable PET, thus reducing the risk of critical encounters. However, the results also reveal that higher update rates do not guarantee crash-free operation, particularly on lower-end hardware. Two factors explain this limitation: (\textit{i}) when excessive amounts of collaborative perception information are received, limited processing resources introduce latency, delaying reactions; and (\textit{ii}) under heavy communication load, dropped packets may create random gaps in the perceived environment. These effects are exacerbated for low- and mid-end hardware, demonstrating that CPS safety performance is not fully scalable with computational capacity. In other words, increasing frequency improves awareness but simultaneously stresses processing units, which can paradoxically reintroduce safety risks.

In contrast, the proposed Context-Aware Adaptive Filter effectively mitigates these issues. By prioritizing contextually relevant and safety-critical information, the adaptive mechanism prevents information overload and reduces processing burden, ensuring consistent situational awareness even on low-end hardware. As shown in \autoref{fig:safety}, the adaptive filtering approach achieves TTC and PET distributions comparable to high-frequency CPS, without crash cases and with improved stability across all hardware tiers.

\paragraph{Decision-making efficiency}

\begin{table}[t]
\centering
\caption{Decision-making efficiency for low-end hardware and (high-end hardware).}
\label{tab:decision_efficiency}
\begin{adjustbox}{width=\linewidth}
\scriptsize
\setlength{\tabcolsep}{4pt}
\renewcommand{\arraystretch}{1.35}
\begin{tabular}{lccccc}
\hline
\multirow{2}{*}{\textbf{Experience}} &
\multirow{2}{*}{\begin{tabular}[c]{@{}c@{}}\textbf{Max}\\\textbf{Queue}\end{tabular}} &
\multirow{2}{*}{\begin{tabular}[c]{@{}c@{}}\textbf{Info}\\\textbf{Dropped}\end{tabular}} &
\multicolumn{2}{c}{\textbf{Average AoI (ms)}} \\
\cline{4-5}
 & & & \textbf{LS + Remote} & \textbf{Remote} \\
\hline
CPS--1\,Hz      & 140.15 (56.2)   & \textbf{0\%} (\textbf{0\%})   & 276 (79)       & 1108 (147) \\
CPS--5\,Hz      & 992.97 (148)    & 67\% (\textbf{0\%})          & 276.5 (63.5)   & 956.1 (80.3) \\
CPS--10\,Hz     & 1000 (256.37)   & 79.8\% (\textbf{0\%})        & \textbf{101} (\textbf{47.6}) & \textbf{442.2} (\textbf{68.9}) \\
Adaptive Filter & \textbf{41.6} (\textbf{24}) & \textbf{0\%} (\textbf{0\%}) & 114.9 (79.1)   & 486.7 (376.8) \\
\hline
\end{tabular}
\end{adjustbox}
\end{table}

\autoref{tab:decision_efficiency} complements the safety analysis by illustrating how hardware capacity impacts decision-making efficiency through queue occupancy, packet loss, and information freshness. The \textit{Max Queue} denotes the average maximum buffer length of \textit{Perceived Object} elements extracted from Cooperative Perception Messages (CPMs). Since queues are capped at $1000$ \textit{Perceived Objects}, values approaching this limit indicate saturation, which inevitably results in packet drops.

The table further reports the \textit{Age of Information} (AoI), defined as the time elapsed between the generation of a data item and its use in the vehicle’s decision-making process. Lower AoI values correspond to more up-to-date information. Local Sensor (LS) data originates from onboard sensing, while \textit{Remote} information is received from neighboring agents via CPMs. At 1~Hz, both hardware tiers process information reliably; however, the average AoI for low-end hardware ($276$~ms) is more than three times higher than that of high-end hardware ($79$~ms), indicating delayed responsiveness on resource-constrained devices.

At higher communication rates (5~Hz and 10~Hz), low-end hardware becomes overloaded: queues saturate and up to $80\%$ of messages are dropped, leading to information gaps and unstable safety margins, as reflected in \autoref{fig:safety}. In contrast, high-end devices maintain stable queues with zero packet loss, keeping AoI below $48$~ms even at 10~Hz. These results demonstrate that conventional CPS configurations do not scale to heterogeneous fleets, as improvements tailored to high-end platforms can destabilize low-end systems.

The proposed \textit{Adaptive Filter} mitigates this mismatch by reducing queue occupancy and preventing packet drops across all hardware tiers. Although low-end devices still exhibit higher average AoI (approximately $115$~ms), safety is preserved through selective filtering and prioritization of decision-critical information. This prioritization stabilizes CPM-derived AoI, yielding a more consistent and reliable information stream. Consequently, the proposed approach alleviates information overload while ensuring that delayed updates remain safety-relevant, enabling scalable and efficient decision-making in heterogeneous CPS deployments.


\paragraph{Communication Benefits}

\begin{table}[h!]
\centering
\renewcommand{\arraystretch}{1} 
\setlength{\tabcolsep}{8pt} 
\caption{Number of packets sent and total transmission sizes for different experience configurations.}
\label{tab:packets_transmissions}
\begin{adjustbox}{width=1\linewidth}
\begin{tabular}{llcc}
\toprule
\textbf{Experience} & \textbf{Config} & \textbf{Packets Sent} & \textbf{Transmission Size (kB)} \\
\midrule
\multirow{3}{*}{CPS (All tiers)} 
 & 1Hz  & 4074  & 1059.66 \\
 & 5Hz  & 20457 & 5328.18 \\
 & 10Hz & 40954 & 10669.81 \\
\midrule
\multirow{3}{*}{Adaptive Filter} 
 & Low-End  & 5638  & 771.68 \\
 & Mid-End  & \textbf{3227} & \textbf{441.64} \\
 & High-End & 3507  & 483.18 \\
\bottomrule
\end{tabular}
\end{adjustbox}
\end{table}

Table~\ref{tab:packets_transmissions} presents the communication overhead of conventional CPS in comparison with the proposed Adaptive Filter. As anticipated, the overhead of traditional CPS grows almost linearly with transmission frequency: at 10~Hz, more than $40{,}000$ packets are transmitted, amounting to over $10$~MB of data. This volume of traffic can quickly saturate the communication channel, increasing the likelihood of congestion and overwhelming receivers with redundant information. 

In contrast, the Adaptive Filter significantly reduces both the number of packets and the total transmission size across all hardware tiers. Packet counts remain between $3{,}200$ and $5{,}600$, with less than $800$~kB transmitted in total—representing more than an 93\% reduction compared to CPS-10Hz. Crucially, this reduction does not come at the expense of safety: as shown earlier, TTC and PET distributions remain comparable to the highest-frequency CPS, while decision-making efficiency improves due to the absence of queue saturation. This demonstrates that context-aware adaptive filtering not only reduces communication load but also preserves the timeliness and usability of critical updates by preventing information overload at the receiver side.



\section{Conclusion}
This paper introduced a frugal communication framework for Collective Perception Service that prioritizes safety-relevant information through context-aware adaptivity. Conventional CPS improves situational awareness but relies on high-frequency transmissions that generate substantial communication overhead, induce latency through overloaded queues, and fail to scale across heterogeneous hardware platforms. We addressed this limitation by designing a Context-Aware Adaptive Filter that dynamically regulates the content and frequency of CPMs according to situational relevance.

Our end-to-end SUMO–Artery simulations show that the proposed approach achieves safety performance comparable to the highest-frequency CPS baseline (10~Hz), while reducing communication volume by more than 93\% and preventing queue saturation. These results demonstrate that context-aware adaptivity enables scalable and reliable collective perception without sacrificing safety, making it well suited for realistic C-ITS deployments.

Overall, the results demonstrate that context-aware adaptivity redefines the trade-off in collaborative perception: instead of choosing between higher safety or lower communication cost, both can be achieved simultaneously. Future work will explore real-world validation on embedded platforms, integration with dynamic network management, and extending prioritization to broader cooperative driving tasks beyond safety-critical perception.

\bibliographystyle{IEEEtran}
\bibliography{biblio}

\end{document}